# Deep Learning-Based Hurricane Resilient Co-planning of Transmission Lines, Battery Energy Storages and Wind Farms

Mojtaba Moradi-Sepahvand, Turaj Amraee, *Senior Member, IEEE, and* Saleh Sadeghi Gougheri

*Abstract*—In this paper, a multi-stage model for expansion co-planning of transmission lines, Battery Energy Storages (BESs), and Wind Farms (WFs) is presented considering resilience against extreme weather events. In addition to High Voltage Alternating Current (HVAC) lines, Multi-Terminal Voltage Source Converter (MTVSC) based High Voltage Direct Current (HVDC) lines are planned to reduce the impact of high-risk events. To evaluate the system resilience against hurricanes, probable hurricane speed (HS) scenarios are generated using Monte Carlo Simulation (MCS). The Fragility Curve (FC) concept is utilized for calculating the failure probability of lines due to extreme hurricanes. Based on each hurricane damage, the probable scenarios are incorporated in the proposed model. Renewable Portfolio Standard (RPS) policy is modeled to integrate high penetration of WFs. To deal with the wind power and load demand uncertainties, a Chronological Time-Period Clustering (CTPC) algorithm is introduced for extracting representative hours in each planning stage. A deep learning approach based on Bi-directional Long Short-Term Memory (B-LSTM) networks is presented to forecast the yearly peak loads. The Mixed-Integer Linear Programming (MILP) formulation of the proposed model is solved using a Benders Decomposition (BD) algorithm. A modified IEEE RTS test system is used to evaluate the proposed model effectiveness.

## Nomenclature

**Indices & Sets**

| | |
|---|---|
| $s, \Omega_S, S, g$ | Index, Set, and total planning stages, Index of time interval. |
| $h, \Omega_H, H$ | Index, Set, and total representative hours. |
| $i, \Omega_B, \Omega_G$ | Index of bus, Sets of buses and conventional units. |
| $\Omega_{sb}, \Omega_w$ | Sets of candidate buses for installing BESs and WFs. |
| $l, L$ | Index of all lines, and total number of transmission lines. |
| $\Omega_{nl}^{ac}, \Omega_{nc}^{ac}$ | Sets of all HVAC candidate lines and HVAC candidate lines in new corridors as a subset of $\Omega_{nl}^{ac}$. |
| $\Omega_{el}, \Omega_{nl}^{dc}$ | Sets of existing lines, and HVDC candidate lines. |
| $v, \Omega_V$ | Index and Set of all Voltage Source Converters. |
| $c, \Omega_C$ | Index and Set of the corridors. |
| $p, \Omega_P, P$ | Index, Set, and total number of linear segments of the conventional units generation cost function. |
| $n, \Omega_N, N$ | Index, Set, and total piecewise linearization blocks. |
| $t, \Omega_T, T$ | Index, Set, and total number of towers for each line. |
| $d, j, \Omega_J, J$ | Index of probable failure scenarios, Index, Set, and total number of most probable failure scenarios due to hurricane. |
| $hs, \Omega_{HS}, \Omega_{HZ}$ | Index and Set of probable hurricane speed scenarios, Set of lines in the hurricane zone. |

Mojtaba Moradi-Sepahvand, Turaj Amraee, and Saleh Sadeghi Gougheri are with the Faculty of Electrical Engineering, K.N. Toosi University of Technology, Tehran, Iran. (e-mails: mojtaba.moradi@email.kntu.ac.ir, amraee@kntu.ac.ir, and salehsadeghi@email.kntu.ac.ir).

**Parameters**

| | |
|---|---|
| $r$, LT | Interest rate and Lifetime of equipment. |
| $IC_l$, $LL_l$ | Investment cost of new line $l$ including the cost of conductors of single/double circuits and towers ($10^6$ $/km$), Line length of all new lines ($Km$). |
| $ICw_i$ | Investment cost of new installed WF ($10^6$ $/MW$). |
| $Rw_l^{ac}, Rw_l^{dc}$ | Right of Way cost for HVAC and HVDC line $l$ including land cost ($10^6$ $/Km$). |
| $AS_l, Vc_v$ | New HVAC substation cost ($10^6$ $), VSCs cost ($10^6$ $/MW$). |
| $Cg_i^p$ | Cost of power generation in each segment $p$ for each conventional unit $i$ ($/MWh$). |
| $Cls_i$, $Cwc_i$ | Load shedding and wind curtailment cost ($/MWh$) in bus $i$. |
| $RU_i, RD_i$ | Ramp up and down limits of conventional unit $i$ ($MW/h$). |
| $\alpha, \beta$ | Expected contribution of WF in supplying the total load until the end of planning horizon, and the maximum annual wind curtailment in the system. |
| $\gamma, \Phi$ | Maximum allowable hourly load shedding in each bus, and annual load shedding in the system, as percentages of each bus load, and the total load. |
| $\vartheta$ | Energy to power ratio of BES. |
| $\{\bullet\}^{max}, \{\bullet\}^{min}$ | Maximum/minimum limits of variables. |
| $Kl, Kb$ | Connectivity matrices of HVDC lines and buses with VSCs. |
| $A, K$ | Directional connectivity matrices of existing and HVAC lines with buses. |
| $Cs_i, Cc_i, dc_i$ | Investment cost of energy capacity ($/MWh$), and power capacity ($/MW$) of each BES in bus $i$. |
| $\eta_c, \eta_d$ | Charging and discharging efficiency of BES. |
| $Wf_h$, $Lf_h$ | Hourly representative wind power and load demand per-unit factors extracted by CTPC algorithm. |
| $Ld_i^{PK}, Lg_s$ | Peak load of bus $i$ ($MW$), and Load Growth factor at stage $s$. |
| $\rho_h, \xi$ | The weight of extracted representative hour $h$, and Reserve cost factor as a percentage of the cost of power generation. |
| $M, \Psi, B$ | Big-M, Base power of the system ($MVA$), Per-unit susceptance of HVAC lines. |
| $\phi, \psi, \chi$ | VSC power loss function coefficients. |
| $SPv_{n,s,v,l,h}$ | Slope of the $n^{th}$ discretization of VSC power capacity. |
| $HD, FP, Hp_{hs}$ | Horizontal distance of towers, Failure probability function of line and tower based on the fragility curve, Probability of the hurricane with speed $hs$. |
| $ni, nh, no, rd$ | Total number of input data, hidden units, and outputs, Dimension of each input data sequence in B-LSTM. |

**Variables**

| | |
|---|---|
| $Z$ | Total Planning Cost (TPC). |
| $Y_{s,l,c}, Yd_{s,l,c}$ | Binary variables of candidate HVAC and HVDC line $l$ at stage $s$ and corridor $c$ (equals 1 if the candidate line is constructed and 0 otherwise). |
| $I_{s,i,h}, U_{s,i,h}$ | On/Off state of conventional units, Charging/Discharging state of BES, in bus $i$ at stage $s$, and hour $h$. |
| $X_{d,l}$ | State of each vulnerable line $l$ in probable failure scenario $d$. |
| $Pe_{s,l,h}$ | Power flow across existing line $l$ at stage $s$ and hour $h$ ($MW$). |
| $Pv_{s,v,l,h}$ | Power capacity of MTVSC $v$ for HVDC line $l$, at stage $s$, and hour $h$ ($MW$). |
| $Pv_{s,v,l,h}^+, Pv_{s,v,l,h}^-$ | Auxiliary positive variables to calculate $\|Pv_{s,v,l,h}\|$. |
| $\Delta_{n,s,v,l,h}$ | Value of the $n^{th}$ block associated with discretization of VSC power capacity ($MW$). |
| $S_{s,i}, C_{s,i}$ | BES energy ($MWh$) and power ($MW$) capacity at stage $s$. |
| $P_{s,i,h}, Ps_{s,i,h,p}$ | Power generation of conventional unit *i*, and segment *p* of unit *i* at stage *s*, and hour *h* (*MW*). |
| $Pw_{s,i}, PC_{s,i,h}$ | Total power capacity of WF *i* at stage *s* (*MW*), and wind curtailment of WF *i* at stage *s* and hour *h* (*MW*). |
| $R_{s,i,h}, LS_{s,i,h}$ | Reserve of conventional unit *i* at stage *s* and hour *h* (*MW*), and load shedding in bus *i* at stage *s* and hour *h* (*MW*) |
| $E_{s,i,h}$ | Stored energy (*MWh*) of BES in bus *i* at stage *s* and hour *h*. |
| $Pd_{s,i,h}, Pc_{s,i,h}$ | Discharging and charging power of BES in bus *i* at stage *s* and hour *h* (*MW*). |
| $Pl_{s,l,c,h}$ | Power flow across new constructed HVAC line *l*, in corridor *c*, at stage *s* and hour *h* (*MW*). |
| $\theta_{s,i,h}$ | Voltage angle of bus *i* at stage *s*, and hour *h*. |
| $bo, bh^{(b)}, bh^{(f)}$ | Bias vector for output gate, backward, and forward hidden layer in B-LSTM. |
| $Whh^{(b)}, Whh^{(f)}$ $Wxh^{(b)}, Wxh^{(f)}$ $Wo$ | Weight vector of backward and forward layers output data, backward and forward layers input data, and the output layer in B-LSTM. |

**Compact Representation**

| | |
|---|---|
| **Y** | Vector of binary decision variables. |
| **S** | Vector of BES power and energy capacity variables. |
| **W** | Vector of WF power capacity variables. |
| **P** | Vector of positive continuous operational variables. |
| **Q** | Vector of free continuous variables. |

I. INTRODUCTION

*A. Motivation and Background*

DISASTERS are the High Impact Low Probability (HILP) events, like hurricanes and earthquakes, affecting the main infrastructures such as power systems. Among HILP disasters, in recent years, the probability of extreme weather events is increased. The power system capability to recover quickly from such events and reduce the impacts is defined as the system resilience [1]. The focus of resiliency is on the rare extreme events that can cause cascading outages, or the outage of a major part of the network, while the conventional reliability or security criteria focus on the most probable contingencies. One of the most effective procedures to improve the power grid resilience against HILP weather events is constructing new transmission lines to increase the power transfer from the generation side to the consumers. The long-distance transmission lines with many towers are vulnerable to hurricanes, while this critical issue is ignored in many Transmission Expansion Planning (TEP) problems [2]. To enhance the transmission system resilience against extreme weather events, in addition to the construction of new HVAC transmission lines, other modern tools such as Battery Energy Storage (BES) devices, Renewable Energy Sources (RESs), and HVDC transmission lines can be taken into account.

A strategy for boosting the system resiliency using the Fragility Curve (FC) and considering a defensive islanding method to mitigate the cascading outages due to weather events is proposed in [3]. A scenario-based methodology to proactively minimize the hurricane effects, e.g., load shedding and generation cost, is addressed in [4]. In [5], to enhance the resilience against disasters, the operation of jointly regional and district integrated energy system is proposed. In [6], a vulnerability constrained generation and transmission expansion model considering the impacts of seismic and intentional events is developed. A planning-oriented model to evaluate the transmission system resilience against typhoon disasters is presented in [2]. The research works in [2-7] rely on HVAC lines to promote the system resilience against disasters. The rapid controllability and flexibility of Multi-Terminal Voltage Source Converter (MTVSC) based HVDC lines power flow can decrease the adjacent HVAC lines loading and enhance bulk power system resilience, since the HVDC lines have high capacity with significant overload capability. Flexible operation of HVDC lines offers several improvements to overall bulk power delivery resilience, such as facilitating the renewable integration and providing control strategies for stability improvement. Also, an MTVSC based HVDC line controls both active and reactive powers through controlling switches at a same time [8]. Additionally, the black-start capabilities of HVDC lines provided by VSC, can reduce the risk of cascading outages [8, 9]. Moreover, in comparison to HVAC lines, high power transfer capability and narrower Right of Way (RoW) are other attractive advantages of HVDC lines [9-11]. In [11], an economic assessment of HVDC lines for increasing the power exchange between eastern and western parts of the US is presented in which resilience enhancement against large scale disruptions is concluded. The cybersecurity of HVDC stations against false data injection attacks is addressed in [9]. More details about promoting the power system resilience using HVDC lines is available in [8].

According to the Renewable Portfolio Standard (RPS) policy, the share of RESs in supplying the system total load demand must be increased. The optimal sitting and sizing of Wind Farms (WFs) should be incorporated in the TEP model to achieve an RPS constrained TEP model. The WFs penetration can increase the power system resilience against the HILP events. In fact, WFs can provide a diversified portfolio of produced power, especially under the conditions that fuel and water transport infrastructures of the conventional power generation units are disrupted due to extreme weather events [12]. Additionally, due to the inherent uncertainty of WFs, it is more critical to consider the resilience of power system in the expansion planning. In this regard, considering flexible ramp reserve of conventional generating units and proper methods for capturing the short-term and long-term uncertainties is essential. Moreover, to handle the intermittency of WFs, reduce the wind curtailment, and also to relieve the congestion of transmission lines, fast responsive BESs are useful devices [13]. In comparison to large conventional units, BESs are capable of supplying power as distributed generation resources during disasters. Besides, BESs facilitate the black start process for the power system [14, 15]. The minimization of energy not supplied using BESs is addressed in [16]. In [14], to enhance the power system resilience under HILP events, the optimal placement and sizing of photovoltaic and BESs in the transmission system are considered. In [17] and [18], expansion models are proposed for co-planning transmission lines and ESs concerning high penetration of WFs without dealing with resiliency. In [18], the possibility of transmission line switching is also considered. In [19], transmission and ES are planned to minimize the investment cost with high WF penetration. In [19], long and short term uncertainties are modeled in a coordinated investment of transmission and ESs. In the previous studies (i.e., [17-19]), the power system resilience against HILP events is not investigated.

An essential challenge in the long-term planning models, especially considering extreme weather events like hurricanes, is dealing with operational uncertainties. The uncertainty of load demand, WFs output power, and yearly load growth need to be captured using efficient methods [20]. One method is considering many scenarios as representatives, which can cause more computational burden. The common techniques to extract limited representative days are K-means (e.g., [19, 21]) and hierarchical clustering (e.g., [20]). These techniques are not able to select representative hours accurately [22]. A Chronological Time-Period Clustering (CTPC) is proposed in [22] to choose the representative hours by considering the varying parameters chronology. In a long-term horizon, the method efficiency in [22] is reduced, especially in capturing peak load values.

### B. Research Gaps and Contributions

In most previous works, either the TEP study is conducted without considering the power system resilience against disasters, or in the resilient models the planning of WFs under RPS policy considering the sitting and sizing of BES devices, along with modeling flexible ramp reserve to improve the power system resilience against disasters are ignored. Besides, the key role of MTVSC based HVDC lines in enhancing the TEP resilience and reducing the risk of cascading outages is not investigated. Other major gaps in the previous resilient TEP models are, 1) The lack of an efficient clustering approach for extracting representative hours in long-term co-planning models by considering the chronology of the time-dependent parameters to capture the operational uncertainties of WFs output power and load demand, 2) The lack of a valid approach to forecast the yearly load growth according to the historical data in long-term co-planning models. Regarding these gaps, the main contributions of this work are summarized as follows.

1) A multi-stage expansion co-planning model for HVAC/DC lines, BES devices, and WFs is proposed to promote the power system resilience against extreme weather events. The high shares of WFs under RPS policy, the sitting and sizing of BES devices, and also modeling flexible ramp reserve for handling the WFs output power and load demand uncertainties are integrated in the proposed TEP model to improve the power system resilience against probable hurricanes. Moreover, the MTVSC based HVDC line is considered as an effective tool to reduce the impact of extreme hurricanes, and facilitate the integration of remote WFs. The FC is utilized to calculate the failure probability of lines due to probable Hurricane Speed (HS) scenarios obtained using Monte Carlo Simulation (MCS).

2) To deal with WFs output power and load demand uncertainties, a modified CTPC algorithm is introduced to extract proper representative hours for each planning stage. The proposed CTPC algorithm is not only capable of handling wind fluctuations, but the yearly peak and valley values of load data can be more accurately captured. Also, a new version of deep Long Short-Term Memory (LSTM) networks with bi-directional memory known as Bi-directional LSTM (B-LSTM) is developed to forecast the yearly load growth.

3) The proposed co-planning model is formulated as a Mixed-Integer Linear Programming (MILP) problem and is solved using a Benders Decomposition (BD) algorithm. The developed BD algorithm facilitates the simultaneous optimization of investment and operational variables by decomposing the main problem to a master problem and several dual sub problems.

## II. FORMULATIONS

The proposed Hurricane Resilient (HR) co-planning model is formulated as an MILP problem based on DC Optimal Power Flow (DC-OPF). The formulations are presented in both general and BD forms. The objective function and the corresponding constraints are introduced as follows.

### 1) Objective Function

The objective function in (1), minimizes the Discounted Present Value (DPV) of investment and operation costs.

$$Min \ Z = \sum_{s \in \Omega_S} \left[ \left( \frac{2}{(1+r)^{2s-1}} \right) \cdot (AL_s + DL_s + BE_s + WF_s) \right] + \sum_{s \in \Omega_S} \left[ \left( \frac{2}{(1+r)^{2s}} \right) \cdot \left[ \sum_{h \in \Omega_H} \rho_h \cdot (GF_{s,h} + LWC_{s,h}) \right] \right] \quad (1)$$

In the proposed multi-stage model, each planning stage is considered 2-years. The DPV for the investment cost is assumed at the beginning of each planning stage. The operational costs are assumed at the end of each stage. In the objective function, the first term represents the Total Investment Cost (TIC), including the cost of HVAC lines ($AL_s$), MTVSC-based HVDC lines ($DL_s$), BESs ($BE_s$), and WFs ($WF_s$). The Total Operating Cost (TOC) is the second term of (1). It includes the linearized cost of generators and the expected flexible ramp reserve cost ($GF_{s,h}$), and the load and wind curtailment costs ($LWC_{s,h}$).

Based on equipment lifetime and Capital Recovery Factor (CRF), the DPV investment cost is converted to Equivalent Annual Cost (EAC) [23]. The EAC of HVAC lines is expressed as (1a). The RoW cost is computed for all new corridors. The substation cost is considered just for the first new corridor.

$$AL_s = \frac{r(1+r)^{LT_l}}{(1+r)^{LT_l}-1} \times \left[ \sum_{l \in \Omega_{nl}^{ac}} [IC_l \cdot LL_l \times (\sum_{c \in \Omega_C} Y_{s,l,c})] + \sum_{l \in \Omega_{nl}^{ac}} [Rw_l^{ac} \cdot LL_l \times (\sum_{c \in \Omega_C} Y_{s,l,c})] + \sum_{l \in \Omega_{nc}^{ac}} [Y_{s,l,c=1} \cdot (AS_l)] \right] \quad (1a)$$

The EAC for HVDC corridors, including the cost of lines, RoW, and the MTVSCs is formulated as (1b).

$$DL_s = \frac{r(1+r)^{LT_l}}{(1+r)^{LT_l}-1} \times \left[ \sum_{l \in \Omega_{nl}^{dc}} [IC_l \cdot LL_\ell \times (\sum_{c \in \Omega_C} Yd_{s,l,c})] + \sum_{l \in \Omega_{nl}^{dc}} [Rw_l^{dc} \cdot LL_l \times (\sum_{c \in \Omega_C} Yd_{s,l,c})] + \sum_{v \in \Omega_V} \sum_{l \in \Omega_{nl}^{dc}} [Kl_l^v \cdot P_l^{max} \cdot Vc_v \cdot (\sum_{c \in \Omega_C} Yd_{s,l,c})] \right] \quad (1b)$$

The EAC of BESs and WFs investment is formulated in (1c) and (1d), respectively.

$$BE_s = \frac{r(1+r)^{LT_{BES}}}{(1+r)^{LT_{BES}}-1} \times \left[ \sum_{i \in \Omega_{sb}} [Cs_i \cdot (S_{s,i}) + Cc_i \cdot (C_{s,i})] \right] \quad (1c)$$

$$WF_s = \frac{r(1+r)^{LT_{WF}}}{(1+r)^{LT_{WF}}-1} \times \left[ \sum_{i \in \Omega_W} ICw_i \cdot Pw_{s,i} \right] \quad (1d)$$

In (1e) the operation cost of $GF_{s,h}$ is presented.

$$GF_{s,h} = \sum_{i \in \Omega_G} \left[ [Cg_i^{p=1} \cdot (P_i^{min} \cdot I_{s,i,h} + \xi \cdot R_{s,i,h})] + \sum_{p \in \Omega_P} [Cg_i^p \cdot Ps_{s,i,h,p}] \right] \quad (1e)$$

The cost of load and wind curtailment is introduced in (1f):

$$LWC_{s,h} = \sum_{i \in \Omega_B} Cls_i \cdot LS_{s,i,h} + \sum_{i \in \Omega_w} Cwc_i \cdot PC_{s,i,h} \quad (1f)$$

### 2) Technical Constraints of the proposed model

The technical constraints of generating units are presented in

(2) to (5). The constraint in (2) limits the minimum and maximum output power of conventional units. The cost function of conventional units is linearized using constraints (3) and (4). In (3), the units hourly output power is defined as a summation of the minimum power and a set of linear generation power segments. In (4), each segment generated power is bounded. The units ramp up and ramp down are defined in (5).

$$P_i^{min}.I_{s,i,h} \leq P_{s,i,h} \leq P_i^{max}.I_{s,i,h} \quad \forall i \in \Omega_G, s \in \Omega_S, h \in \Omega_H \quad (2)$$

$$P_{s,i,h} = P_i^{min}.I_{s,i,h} + \sum_{p=1}^{P} Ps_{s,i,h,p} \quad \forall i \in \Omega_G, s \in \Omega_S, h \in \Omega_H \quad (3)$$

$$0 \leq Ps_{s,i,h,p} \leq (P_i^{max} - P_i^{min}).I_{s,i,h}/P \quad \forall i \in \Omega_G, s \in \Omega_S, h \in \Omega_H, p \in \Omega_P \quad (4)$$

$$\begin{cases} P_{s,i,h} - P_{s,i,h-1} \leq RU_i & \forall i \in \Omega_G, s \in \Omega_S, h \in \Omega_H \\ P_{s,i,h-1} - P_{s,i,h} \leq RD_i & \forall i \in \Omega_G, s \in \Omega_S, h \in \Omega_H \end{cases} \quad (5)$$

The constraints given by (6) and (7) are defined to deal with the RPS policy. The bounds of the total power capacity of WFs are described in (6). The minimum installed capacity of WFs for supplying the load demand at each 2-years planning stage is considered as a certain percentage of the peak load. According to (7), it is assumed that at the end of the planning horizon, the total installed capacity of WFs is at least $\alpha\%$ of the peak load.

$$0 \leq Pw_{s,i} \leq Pw_i^{max} \quad \forall i \in \Omega_w, s \in \Omega_S \quad (6)$$

$$[\alpha \times s/S] \times (1 + Lg_s).\sum_{i \in \Omega_B} Ld_i^{pk} \leq \sum_{i \in \Omega_w} Pw_{s,i} \quad \forall s \in \Omega_S \quad (7)$$

In the presence of WFs and BESs, considering the wind power curtailment and load shedding is necessary. The limits of hourly wind curtailment for each WF are defined based on (8). The maximum permitted annual wind curtailment is defined in (9) as a certain percentage of the total expected power output of WFs. The permitted hourly load shedding in each bus is bounded using (10). Based on (10), the upper bound of load shedding is a certain percentage of the expected hourly load. In (11), the maximum allowable annual load shedding is assumed as a certain percentage of the annual expected load.

$$0 \leq PC_{s,i,h} \leq Wf_h.Pw_{s,i} \quad \forall i \in \Omega_w, s \in \Omega_S, h \in \Omega_H \quad (8)$$

$$\sum_{i \in \Omega_w}\sum_{h \in \Omega_H} PC_{s,i,h} \leq \beta \times \sum_{i \in \Omega_w}\sum_{h \in \Omega_H} Wf_h.Pw_{s,i} \quad \forall s \in \Omega_S \quad (9)$$

$$0 \leq LS_{s,i,h} \leq \gamma.(1 + Lg_s).Lf_h.Ld_i^{pk} \quad \forall i \in \Omega_B, s \in \Omega_S, h \in \Omega_H \quad (10)$$

$$\sum_{i \in \Omega_B}\sum_{h \in \Omega_H} LS_{s,i,h} \leq \Phi \times (1 + Lg_s).\sum_{i \in \Omega_B}\sum_{h \in \Omega_H} Lf_h.Ld_i^{pk} \quad \forall s \in \Omega_S \quad (11)$$

As given in (12) to (14), the flexible ramp reserve is considered in the proposed model to more accurately capture the uncertainty of load and WFs. The reserve limits are enforced by (12). In (13), each conventional unit maximum capacity is defined as its reserve plus the output power. According to (14), the total hourly reserve is assumed to be at least 5% and 3% of the hourly expected output power of WFs, and the total load [24].

$$0 \leq R_{s,i,h} \leq P_{s,i,h} \quad \forall i \in \Omega_G, s \in \Omega_S, h \in \Omega_H \quad (12)$$

$$R_{s,i,h} + P_{s,i,h} \leq P_i^{max} \quad \forall i \in \Omega_G, s \in \Omega_S, h \in \Omega_H \quad (13)$$

$$\sum_{i \in \Omega_G} R_{s,i,h} \geq (5\%) \times \sum_{i \in \Omega_w} Wf_h.Pw_{s,i} + (3\%) \times (1 + Lg_s).Lf_h.\sum_{i \in \Omega_B} Ld_i^{pk} \quad \forall s \in \Omega_S, h \in \Omega_H \quad (14)$$

To model BES devices, the constraints in (15) to (23) are presented. The hourly charging and discharging power of BES are bounded using (15) and (16). The BES hourly state of charging or discharging is determined by (17) and (18). In (19), the stored energy level in BES is defined as the stored energy at the previous hour plus the energy exchange at the current hour.

At the first hour of each stage, the stored energy at the last hour of previous stage is also modeled in (19). The energy to power ratio of BES is represented in (20). The energy level, and power and energy capacity of BES are bounded using (21) to (23).

$$0 \leq \eta_c.Pc_{s,i,h} \leq C_{s,i} \quad \forall i \in \Omega_{sb}, s \in \Omega_S, h \in \Omega_H \quad (15)$$

$$0 \leq 1/\eta_d.Pd_{s,i,h} \leq C_{s,i} \quad \forall i \in \Omega_{sb}, s \in \Omega_S, h \in \Omega_H \quad (16)$$

$$\eta_c.Pc_{s,i,h} \leq C_i^{max}.U_{s,i,h} \quad \forall i \in \Omega_{sb}, s \in \Omega_S, h \in \Omega_H \quad (17)$$

$$1/\eta_d.Pd_{s,i,h} \leq C_i^{max}.(1 - U_{s,i,h}) \quad \forall i \in \Omega_{sb}, s \in \Omega_S, h \in \Omega_H \quad (18)$$

$$E_{s,i,h} = E_{s,i,h-1} + E_{s-1,i,h+(H-1)} + (\eta_c.Pc_{s,i,h}) - (1/\eta_d.Pd_{s,i,h}) \quad \forall i \in \Omega_{sb}, s \in \Omega_S, h \in \Omega_H \quad (19)$$

$$C_{s,i}.\vartheta \leq S_{s,i} \quad \forall i \in \Omega_{sb}, s \in \Omega_S \quad (20)$$

$$0 \leq E_{s,i,h} \leq S_{s,i} \quad \forall i \in \Omega_{sb}, s \in \Omega_S, h \in \Omega_H \quad (21)$$

$$0 \leq C_{s,i} \leq C_i^{max} \quad \forall i \in \Omega_{sb}, s \in \Omega_S \quad (22)$$

$$0 \leq S_{s,i} \leq S_i^{max} \quad \forall i \in \Omega_{sb}, s \in \Omega_S \quad (23)$$

Each existing line power flow and its limits are considered using (24) and (25), respectively. Each candidate HVAC line power flow and the corresponding limits are given by (26) and (27).

$$Pe_{s,l,h} = \sum_{i \in \Omega_B} \Psi.B_l.A_i^l.\theta_{s,i,h} \quad \forall l \in \Omega_{el}, s \in \Omega_S, h \in \Omega_H \quad (24)$$

$$-P_l^{max} \leq Pe_{s,l,h} \leq P_l^{max} \quad \forall l \in \Omega_{el}, s \in \Omega_S, h \in \Omega_H \quad (25)$$

$$-M_l.(1 - Y_{s,l,c}) \leq Pl_{s,l,c,h} - \sum_{i \in \Omega_B} \Psi.B_l.K_i^l.\theta_{s,i,h} \leq M_l.(1 - Y_{s,l,c}) \quad \forall l \in \Omega_{nl}^{ac}, s \in \Omega_S, c \in \Omega_C, h \in \Omega_H \quad (26)$$

$$-P_l^{max}.Y_{s,l,c} \leq Pl_{s,l,c,h} \leq P_l^{max}.Y_{s,l,c} \quad \forall l \in \Omega_{nl}^{ac}, s \in \Omega_S, c \in \Omega_C, h \in \Omega_H \quad (27)$$

The formulations of MTVSC-based HVDC lines are given in (28) to (34). The VSC power loss (i.e., $P_v^{lss}$), which is the major part of losses in a HVDC system, is presented by (28). The coefficients of $\phi, \psi$, and $\chi$ in (28) represent the constant losses (e.g., filter losses, transformer core losses), the linear losses (e.g., switching losses), and the quadratic losses (e.g., transformer, phase reactor copper losses, and conductor losses), respectively. The quadratic term of VSC loss function is linearized using the piecewise linear method [10], as expressed in (29) to (32). The coupling constraint between two sides of VSC is addressed by (33) for each installed HVDC line. In (33), the linearized losses of both correspond VSCs, are considered. Based on the installed HVDC lines, power capacity of each VSC is restricted by (34).

$$P_v^{lss} = \phi + \psi.Pv_v^{ac\,side} + \chi.(Pv_v^{ac\,side})^2$$
$$(Pv_v^{ac\,side})^2 \approx \sum_{n \in \Omega_N} SPv_{n,s,v,l,h}.\Delta_{n,s,v,l,h} \quad \forall v \in \Omega_V, l \in \Omega_{nl}^{dc}, s \in \Omega_S, h \in \Omega_H \quad (28)$$

$$SPv_{n,s,v,l,h} = (2n - 1).[P_l^{max}]/N \quad \forall n \in \Omega_N, v \in \Omega_V, l \in \Omega_{nl}^{dc}, s \in \Omega_S, h \in \Omega_H \quad (29)$$

$$0 \leq \Delta_{n,s,v,l,h} \leq [P_l^{max}]/N \quad \forall n \in \Omega_N, v \in \Omega_V, l \in \Omega_{nl}^{dc}, s \in \Omega_S, h \in \Omega_H \quad (30)$$

$$\sum_{n \in \Omega_N} \Delta_{n,s,v,l,h} = Pv_{s,v,l,h}^+ + Pv_{s,v,l,h}^- \quad \forall v \in \Omega_V, l \in \Omega_{nl}^{dc}, s \in \Omega_S, h \in \Omega_H \quad (31)$$

$$Pv_{s,v,l,h} = Pv_{s,v,l,h}^+ - Pv_{s,v,l,h}^- \quad \forall v \in \Omega_V, l \in \Omega_{nl}^{dc}, s \in \Omega_S, h \in \Omega_H \quad (32)$$

$$\sum_{v \in \Omega_V} Kl_l^v.[(1 - \psi).Pv_{s,v,l,h} - \chi.(\sum_{n \in \Omega_N} SPv_{n,s,v,l,h}.\Delta_{n,s,v,l,h})] = 2\phi.\sum_{c \in \Omega_C} Yd_{s,l,c} \quad \forall l \in \Omega_{nl}^{dc}, s \in \Omega_S, h \in \Omega_H \quad (33)$$

$$-(Kl_l^v.P_l^{max}.\sum_{c \in \Omega_C} Yd_{s,l,c}) \leq Pv_{s,v,l,h} \leq (Kl_l^v.P_l^{max}.\sum_{c \in \Omega_C} Yd_{s,l,c}) \quad \forall v \in \Omega_V, l \in \Omega_{nl}^{dc}, s \in \Omega_S, h \in \Omega_H \quad (34)$$

The power balance constraint in each bus is defined in (35). The generated power of conventional units and WFs considering wind curtailment, power flow of existing, new HVAC, and new HVDC lines, power exchange of BESs, load demand, and load shedding are all considered in (35).

$$P_{s,i,h} + [Wf_h.Pw_{s,i} - PC_{s,i,h}] + [Pd_{s,i,h} - Pc_{s,i,h}] - \sum_{l \in \Omega_{el}} A_i^l.Pe_{s,l,h} - \sum_{l \in \Omega_{nl}^{ac}}\sum_{c \in \Omega_C} K_i^l.Pl_{s,l,c,h} - \sum_{l \in \Omega_{nl}^{dc}}\sum_{v \in \Omega_V} Kb_i^v.Pv_{s,v,l,h} = ((1 + Lg_s).Lf_h.Ld_i^{PK}) - LS_{s,i,h} \quad \forall i \in \Omega_B, s \in \Omega_S, h \in \Omega_H \quad (35)$$

The proposed formulations considering the probable scenarios to enhance the system resilience against hurricanes is recast in the following to be solved using a BD algorithm.

*3) Benders Decomposition*

In this part, the MILP formulations represented by (1), (1a) to (1f), and (2) to (35), are reformulated in an equivalent compact BD form. The problem is decomposed into one Master Problem (MP) and three Dual Sub-Problems (DSPs). In MP, the optimization of binary decision variables is addressed. In DSPs, the feasibility or optimality of MP solution, WFs and BESs investment cost, and optimization of the system operation are evaluated. An equivalent compact form for the objective function (1), (1a) to (1f), and the constraints (2)-(35) is introduced in the following.

$$\text{Min } I_L^T Y + I_S^T S + I_W^T W + O_C^T P \quad (36)$$

s.t.
$$CW + DP + EQ = F \quad : \sigma \quad (37)$$
$$G_1 Y + H_1 S + J_1 W + K_1 P + L_1 Q = M \quad : \lambda \quad (38)$$
$$G_2 Y + H_2 S + J_2 W + K_2 P + L_2 Q \geq N \quad : \mu \quad (39)$$

$Y \in \{0,1\}, \quad S, W \& P \geq 0, \quad Q: free$
$Y = \{Y, Yd, U, I\}, \quad Q = \{\theta, Pl, Pe, Pv\}$
$P = \{P, Ps, Pv^+, Pv^-, \Delta, R, Pd, Pc, E, PC, LS\}, S = \{\mathcal{S}, C\}, W = \{Pw\}$
$\sigma \& \lambda: free, \quad \mu \geq 0$

The objective function in (36) represents the objective function given by (1), (1a) to (1f). The constraint in (37) denotes (35). The constraints (3), (19), (24), (29), and (31) to (33) are compacted in (38). The constraint in (39) corresponds to the constraints of (2), (4) to (18), (20) to (25), (26), (27), (30), and (34). The dual variables $\sigma$, $\lambda$, and $\mu$ are introduced for (37), (38) and (39), respectively. $I_L$, $I_S$, and $I_W$ are investment cost vectors, and $O_C$ is operation cost vector. The $C, D, E, F, G_1, G_2, H_1, H_2, J_1, J_2, K_1, K_2, L_1, L_2, M$ and $N$ are relevant matrices. For example, in constraint (37), which is the compact form of constraint (35), the matrix $E$ is the compact representative of the coefficients of variable $Q$. Note that $Q$ is compact representative of four free continuous variables, i.e., $Q = \{\theta, Pl, Pe, Pv\}$.

*1) Dual Sub-Problem*

Before introducing DSP, the constraint (40) is added to sub-problem as an auxiliary constraint (with dual variable $\pi$).

$$IY_{sp} = \overline{Y} \quad : \pi \quad I: Identity\ Matrix, \quad \pi: free \quad (40)$$

The binary variables obtained from MP are assumed as constants. The LP formulation of DSP is defined by (41) to (46).

$$\text{Max } F^T \sigma + M^T \lambda + N^T \mu + \overline{Y}^T \pi \quad (41)$$

s.t.
$$D^T \sigma + K_1^T \lambda + K_2^T \mu \leq O_C \quad (42)$$
$$H_1^T \lambda + H_2^T \mu \leq I_S \quad (43)$$
$$C^T \sigma + J_1^T \lambda + J_2^T \mu \leq I_W \quad (44)$$
$$G_1^T \lambda + G_2^T \mu + I\pi \leq 0 \quad (45)$$
$$E^T \sigma + L_1^T \lambda + L_2^T \mu = 0 \quad (46)$$

After considering initial values for $\overline{Y}$, and solving the DSP, if the solution is bounded, the optimality cut for MP is constructed, and the Upper Bound (UB) is calculated as follows.

$$UB = F^T \sigma + M^T \lambda + N^T \mu + \overline{Y}^T \pi + I_L^T \overline{Y} \quad (47)$$

If DSP solution is unbounded, the following Modified DSP (MDSP) is executed to construct the feasibility cut for MP.

*2) Modified DSP*

To deal with unbounded conditions in DSP and remove the extreme rays, an MDSP is defined. Its objective function is assumed as (41), and its constraints are (42) to (46), all with Right-Hand-Side (RHS) equal to zero. Moreover, (48), as an auxiliary constraint, is added.

$$\sigma \leq 1 \quad (48)$$

Another DSP, i.e., Resilience DSP (RDSP), is defined to calculate each resilience scenario risk.

*3) Resilience DSP*

The RDSP is defined similar to DSP, however, limits of load shedding are removed in RDSP. The load shedding limits appear in the objective function (41). After obtaining the MP investment decision variables and evaluating probable failure scenarios, the risk of selected scenarios is assessed using RDSP.

*4) Master Problem*

The IP formulation of MP is represented as follows:

$$\text{Min } Z_{lower} \quad (49)$$

s.t.
$$Z_{lower} \geq I_L^T Y \quad (50)$$
$$Z_{lower} \geq I_L^T Y + [F^T \bar{\sigma} + M^T \bar{\lambda} + N^T \bar{\mu}]^{(v)} + \bar{\pi}^{(v)} \cdot (Y - \overline{Y}^{(v-1)}) \quad (51)$$
$$[M^T \bar{\lambda} + N^T \bar{\mu} + F^T \bar{\sigma}]^{(v)} + \bar{\pi}^{(v)} \cdot (Y - \overline{Y}^{(v-1)}) \leq 0 \quad (52)$$

As given by (49), MP objective function is equal to the Lower Bound (LB) of the problem. The constraint of (50) denotes the investment cost of integer decision variables, and the optimality and feasibility cuts are defined using (51) and (52), respectively. The iteration number is $v$. After solving MP, if the tolerance in (53) is satisfied, the algorithm is terminated.

$$\frac{(UB - LB)}{UB} \leq \varepsilon \quad (53)$$

The mathematical basics and also the required proofs of BD algorithm can be found in [25].

## III. OVERALL STRUCTURE OF THE PROPOSED MODEL

Based on the BD formulation, the overall structure of the proposed model is presented in Fig. 1. The proposed model is formulated as an MILP problem and is decomposed to an MP, containing the binary decision variables, and three DSP (i.e., DSP, MDSP, and RDSP), including the continuous variables. According to Fig. 1, the input data and initial values are first defined, and then the iterative process of BD algorithm is begun.

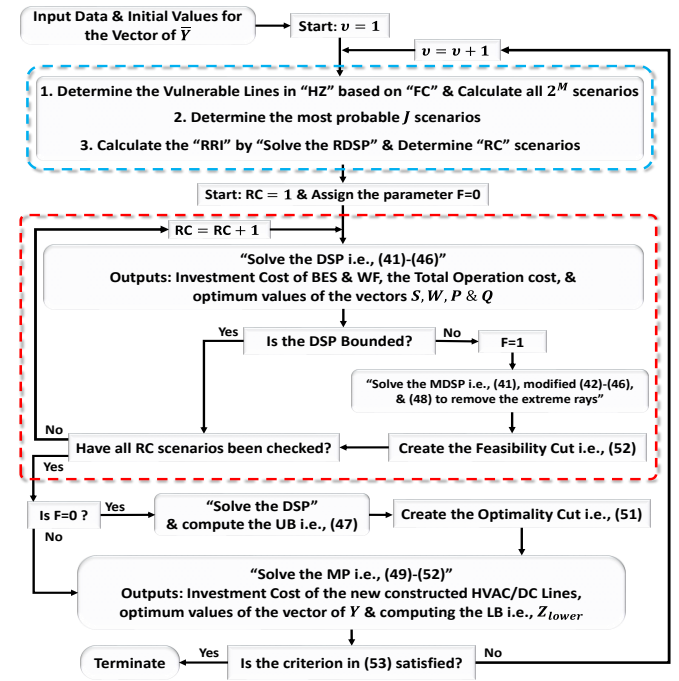

Fig. 1. The Overall Structure of the Proposed Model

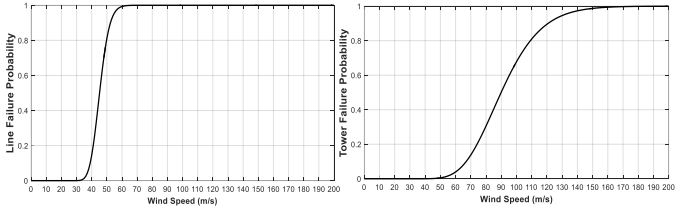

Fig. 2. The Fragility Curves of line (left) and tower (right)

- *Resilience Risk Index*

Regarding the binary decision variables obtained from MP, the system resilience against hurricanes, assuming all existing and new constructed HVAC/DC lines in the considered Hurricane Zone (HZ), is evaluated in each iteration. The process determined by the blue dashed box in Fig. 1, is introduced to distinguish the Resilience Contingencies (RCs) for the proposed HR co-planning model. Using the MCS and Kantorovich distance scenario reduction technique, probable HS scenarios are generated regarding the HS probability distribution function [2]. The hurricane characteristics are assumed homogenous in the HZ. In step 1, to calculate each line hurricane dependent failure probability, the FC illustrated in Fig. 2 [3], is utilized for all lines and towers. Using the FC, the most vulnerable lines (i.e., $M$ lines) due to each HS scenario are recognized according to (54) [2, 3]. A line can be failed due to a tower collapse or a direct line outage. So, the line failure probability is expressed as a function of HS based on the probability of individual line and all correspond towers failure.

$$LF_{l,hs} = FP_{l,hs} + \left[1 - \left(1 - FP_{t,hs}\right)^T\right] - FP_{l,hs} \cdot \left[1 - \left(1 - FP_{t,hs}\right)^T\right] \quad (54)$$
$$T = LL_l/HD \qquad \forall l \in \Omega_{HZ}, hs \in \Omega_{HS}, t \in \Omega_T$$

After selecting $M$ vulnerable lines with $LF > 1\%$ (because the normal failure probability of lines is 1% [3]), configurations of all $2^M$ possible scenarios due to each HS scenario are formed using a simple optimization problem. Then, the probability of each scenario is calculated using (55).

$$Sp_{d,hs} = \prod_{l=1}^{M}[X_{d,l} \cdot LF_{l,hs} + (1 - X_{d,l}) \cdot (1 - LF_{l,hs})] \quad (55)$$
$$\forall hs \in \Omega_{HS}, \ d = 1, \dots, 2^M$$

In step 2, all $2^M$ possible scenarios are reduced to $J$ most probable scenarios with a probability more than $20\% \times \left[\max_{d\&hs}(Sp_{d,hs})\right]$. In step 3, considering line outages in each scenario $j$ under corresponding HS scenario $hs$, the RDSP (i.e., (41)-(46) without load shedding limits) is solved. Then, each scenario impact is calculated as the amount of load shedding. Therefore, in (56) the Resilience Risk Index (RRI) is computed.

$$RRI_{j,hs} = Hp_{hs} \times Sp_{j,hs} \times \sum_{s \in \Omega_S} \sum_{i \in \Omega_B} \sum_{h \in \Omega_H} LS_{s,i,h} \quad \forall j \in \Omega_J, hs \in \Omega_{HS} \quad (56)$$

Assuming normalized RRIs, scenarios with an RRI more than 20% are considered as the chosen RC scenarios.

- *The proposed hurricane resilient Co-planning model*

The introduced iterative process in the red dashed box of Fig. 1, is executed for each RC scenario. The hourly power output of conventional units and WFs, flexible ramp reserve requirement, load shedding and wind curtailment, hourly power exchanges of BESs, power flow of all lines, power capacity of MTVSCs, and the yearly installed capacity of BESs and WFs are optimized in the DSP. As shown in Fig. 1, after solving the DSP for each RC, if the solution is unbounded, the parameter 'F' is set to 1, and the MDSP (i.e., (41), modified (42)-(46) and (48)) is solved to generate a feasibility cut for MP. After evaluating all RCs, if there is no unbounded DSP (i.e., 'F' is equal to 0), the UB is calculated, and an optimality cut is constructed and transferred to MP. In MP, the investment cost of new HVAC/DC lines, and the charging/discharging states of BESs, considering the generated cuts, are optimized. After solving MP and computing the LB, if the criterion defined by (53) is satisfied, the algorithm is terminated; otherwise, the next iteration is begun.

IV. DEEP LEARNING AND CHRONOLOGICAL CLUSTERING

In this paper, a modified CTPC and a deep learning-based B-LSTM approaches are used to select the proper representative hours and yearly load growth forecasting.

- *Proposed Chronological Time-Period Clustering Method*

Among clustering methods, hierarchical clustering is widely used because its output is independent of the initial parameters and can provide various options for merging the clusters. Although the previous methods work well with datasets containing limited data, the accuracy is decreased for large-scale datasets, especially in capturing the peak load values. To solve this deficiency, a CTPC algorithm is proposed in which both representative days and hours are considered. In the proposed CTPC algorithm, the correlation between load and wind data is considered which is important to capture the operation of WFs in the presence of BESs. By considering both representative days and hours, the daily and hourly trends of data in the extracted clusters are captured, that increases the accuracy of the proposed method. Also, this algorithm has a better performance for large scale and high variation datasets. The steps of the proposed CTPC algorithm to select $\mathcal{D}$, and $H$ representative days and hours are given in Fig. 3.

| | |
|---|---|
| Selecting representative days | 1. Set initial number of representative days (ð) equal to the total number of days in dataset. |
| | 2. Calculate the centroid of each cluster $\ell$ ($Cn_\ell = \frac{1}{Dn_\ell}\sum_{e=1}^{Dn_\ell} D_e$), $Dn_\ell$, and $D_e$ are the number of days in cluster $\ell$, and the vector of elements, respectively. |
| | 3. Determine the dissimilarity of each pair of adjacent clusters $\ell$, $\sigma$ ($Ds_{\ell,\sigma} = \sqrt{\frac{2 Dn_\ell Dn_\sigma}{Dn_\ell + Dn_\sigma}} \|Cn_\ell - Cn_\sigma\|_2$). |
| | 4. Merge two closest adjacent clusters ($\hat{\ell}$, $\hat{\sigma}$) based on the dissimilarity matrix. |
| | 5. Update the number of clusters (ð→ð − 1). |
| | 6. If ð = 𝒟 go to the next step, otherwise return to step **2**. |
| | 7. Create a new dataset with reduced size of $\mathcal{T}$ ($\mathcal{T} = \mathcal{D} \times 24$) by putting together the cluster centroids and consider number of days in each cluster as the weights. |
| Selecting representative hours | 1. Assign initial number of new clusters ($\tau$) equal to the total number of hours ($\mathcal{T}$). |
| | 2. Calculate the centroid of each cluster $b$ ($Cn_b = \frac{1}{Hn_b}\sum_{e=1}^{Hn_b} D_e$), $Hn_b$ and $D_e$ are the number of hours in cluster $b$ and the vector of elements, respectively. |
| | 3. Determine the dissimilarity of each pair of adjacent clusters $b$, o ($Ds_{b,o} = \sqrt{\frac{2 Hn_b Hn_o}{Hn_b + Hn_o}} \|Cn_b - Cn_o\|_2$). |
| | 4. Merge two closest adjacent clusters ($\hat{b}$, $\hat{o}$) based on the dissimilarity matrix. |
| | 5. Update the number of clusters ($\tau \to \tau - 1$). |
| | 6. If $\tau = H$ go to the next step, otherwise go to step **2**. |
| | 7. Extract $H$ representative hours as the clusters centroids $Cn_b$. |
| | 8. Consider number of hours in each cluster as final weights. |

Fig. 3. The proposed chronological time-period clustering algorithm

- *B-LSTM Network for Yearly Load Growth Forecasting*

LSTM networks are the modified versions of Recurrent Neural Networks (RNNs), which can overcome the gradient vanishing problem in the RNNs, by various operation gates (input, output, and forget gates). This superiority of LSTM networks results in a significant improvement in forecasting tasks. One-directional LSTM networks neglect the information data that includes future features. To overcome this issue, in this paper, the B-LSTM network with bi-directional memory is utilized to consider the whole temporal horizon, which brings a strong memory to extract useful features and high accuracy. The B-LSTM network is represented as follows [26]:

$$\overrightarrow{HS}_g = tanh(XB_g Wxh^{(f)} + \overrightarrow{HS}_{g-1}) Whh^{(f)} + bh^{(f)} \quad (57)$$

$$\overleftarrow{HS}_g = tanh(XB_g Wxh^{(b)} + \overleftarrow{HS}_{g-1}) Whh^{(b)} + bh^{(b)} \quad (58)$$

$$Ofn = HS_g Wo + bo \quad (59)$$

where $XB_g \epsilon \Re^{ni \times rd}$ is the mini-batch input data at time $g$, $\overrightarrow{HS}_g \epsilon \Re^{ni \times nh}$, $\overleftarrow{HS}_g \epsilon \Re^{ni \times nh}$ are backward and forward hidden layers, and $Ofn \epsilon \Re^{ni \times no}$ is the final output data. The model variables are demonstrated by $Wxh^{(f)} \epsilon \Re^{rd \times nh}, Whh^{(f)} \epsilon \Re^{nh \times nh}, bh^{(f)} \epsilon \Re^{rd \times nh}, Wxh^{(b)} \epsilon \Re^{rd \times nh}, Whh^{(b)} \epsilon \Re^{nh \times nh}, bh^{(b)} \epsilon \Re^{1 \times nh}$. The overall structure of B-LSTM network is illustrated in Fig. 4. In this structure, $O_g$ is the output gate data vector in time $g$. As shown in Fig. 4, at first, the outputs of two hidden layers, i.e., $\overrightarrow{HS}_g$ and $\overleftarrow{HS}_g$, are computed. Then, the final output is calculated based on concatenation of $\overrightarrow{HS}_g$ and $\overleftarrow{HS}_g$. The details of LSTM and related equations are available in [26]. The overall structure of the proposed method for handling uncertainties is presented in Fig. 5.

## V. SIMULATION RESULTS

The simulation results are given in this section. Firstly, the CTPC and B-LSTM methods performance is evaluated. Secondly, the results of the HR co-planning model are analyzed. The MATLAB software is used for simulating the proposed CTPC method [27]. The B-LSTM deep learning method is developed using Phyton software. Finally, the optimization of the proposed HR co-planning model is carried out using CPLEX solver in GAMS [28]. All simulations are executed using a PC with Intel Core i7, 4.2 GHz 7700 CPU, and 32 GB of memory.

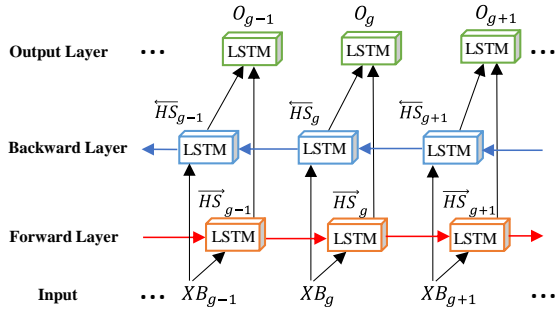

Fig. 4. The overall structure of B-LSTM Network

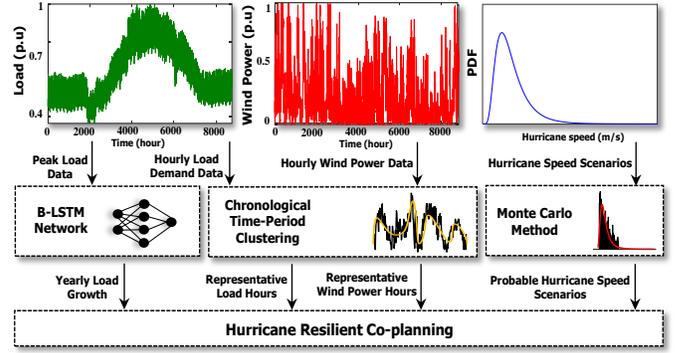

Fig. 5. The structure of the proposed method to handle the uncertainties

- *The CTPC and B-LSTM Network Numerical Results*

The effectiveness of the proposed CTPC is evaluated over load and wind data in 2019, extracted from [29] and [30], respectively. As shown in Fig. 6, firstly, the 365 days of data are reduced to 120 days based on the dissimilarity index. Then, by considering the Normalized Root Mean Square Error (NRMSE) criterion presented in [31] to achieve the proper number of representatives, 96 representative hours are extracted to cover the uncertainties with a tractable complexity. In Table I, to show the performance of proposed CTPC, the obtained results are compared with the algorithm in [22] in terms of the Error Criterion (EC) defined by (60), and capturing peak and valley values of data. The results confirm the superiority of proposed CTPC. In (60), $RD$ and $RH$ are real data and representatives.

$$EC = \frac{1}{F} \sum_{\mathcal{F} \in \Omega_F} \min_{h \in \Omega_H} \{RD_{\mathcal{F}} - RH_h\}, \qquad \Omega_F = 1, \ldots, 8760 \quad (60)$$

To evaluate the B-LSTM network effectiveness in forecasting the yearly load growth, Iran peak load data from 1968 to 2020 [29], are used. The input data are split into 80% and 20% for training and test tasks, L2 regulation, dropout, and mini-batch techniques are used to avoid overfitting and convergence problems. The maximum number of epochs is considered as 1000. To verify the accuracy of B-LSTM network, three well-known error criteria (i.e., Mean Absolute Error (MAE), Mean Absolute Percentage Error (MAPE), and Root Mean Square Error (RMSE)) are utilized for three test years. As given in Table II, the error is compared with two benchmark methods (i.e., LSTM and Multi-Layer Perceptron (MLP)). The considered B-LSTM network presents a better performance for annual peak load forecasting. So, this method is utilized to forecast the yearly peak load over the planning horizon, as presented in Table III.

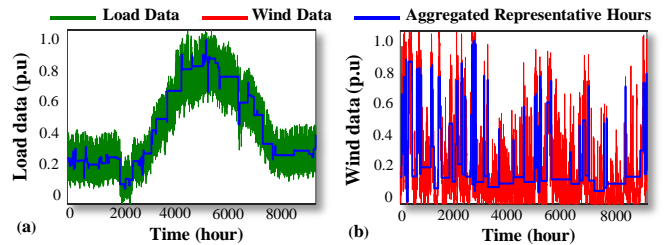

Fig. 6. The proposed CTPC algorithm results for Load (a) and Wind (b) data

- *Results of the Proposed HR Co-Planning Model*

The proposed HR co-planning model performance is verified over a modified IEEE 24-Bus test system. The planning time horizon is assumed to be 8-years (i.e., four 2-years stages). The LT of new HVAC/DC lines, BESs, and WFs are considered as 50, 10, and 20 years, respectively. The construction cost for new HVAC/DC lines and RoW, new HVAC substation, and new MTVSC is assumed as 1/0.96 and 0.034/0.028 $10^6\$/Km$, 3.358 $10^6\$$, and 0.202 $10^6\$/MW$ respectively. For double circuit HVAC lines, the cost of new line, RoW, and substation is 1.6×(1), 1.142×(0.034) $10^6\$/Km$, and 2×(3.358) $10^6\$$, respectively [32]. The $HD$ is assumed to be 500m and the $\phi$, $\psi$, and $\chi$ are considered 0.12, 0.0029, and 0.00031 [33]. BES power and energy capacity costs are assumed to be 500 $\$/Kw$ and 50 $\$/Kwh$ [17]. Both $\eta_c$ and $\eta_d$ are assumed to be 0.9. The construction cost of WF, and load shedding/wind curtailment cost are considered 2 $10^6\$/MW$, and 1000/2000 $\$/Mwh$, respectively. The $\alpha$, $\beta$, and $\vartheta$ are considered 20%, 40%, and $3h$. In the DSP $\gamma$ and $\Phi$ are both zero, while in the RDSP both considered 100%. The $r$ and $\xi$ are 5% and 10%.

The modified IEEE 24-bus test system with 30 existing single circuit and 4 double circuit lines and 11 conventional units can be seen in Fig. 7. Two new buses, i.e., 25 and 26, are candidates to be connected to the system using new HVAC/DC lines. The buses 3, 5, 20, 25 and 26 are BES installation candidate buses with maximum energy and power of 1000 MWh and 250 MW. It is assumed that WFs installation candidate buses are 6, 9, 23, 25, and 26 with maximum capacity of 200, 200, 200, 500 and 500 MW, respectively. Besides, 15 HVAC, and 6 HVDC candidate lines are considered. All required data and parameters of existing and new lines and generators are available in [34].

The uncertainty of HS is considered using five probable HS scenarios generated by MCS and Kantorovich distance scenario reduction technique. To verify the proposed HR co-planning model, four schemes are defined. In **scheme I**, the resilience of the model against hurricanes, BES devices, and HVDC lines are ignored. In **scheme II**, the resilience of the model against hurricanes is ignored, but all planning options are considered. In **scheme III**, the complete model is analyzed considering resilience against hurricanes and ignoring BES devices and HVDC lines. In **scheme IV**, the complete resilience-oriented model is evaluated considering all planning options.

TABLE I
VALIDATION OF THE PROPOSED CTPC RESULTS

| Method | EC (p.u) (i.e., (60)) | | Extremum values of representatives | | | |
|---|---|---|---|---|---|---|
| | Load | Wind | Maximum (p.u) | | Minimum (p.u) | |
| | | | Load | Wind | Load | Wind |
| Proposed CTPC | 0.0044 | 0.0077 | 0.9632 | 0.9430 | 0.4035 | 0.0143 |
| Ref. [22] | 0.0062 | 0.0126 | 0.9147 | 0.9198 | 0.4187 | 0.0435 |

TABLE II
ERROR CRITERIA OF DIFFERENT METHODS FOR TEST YEARS LOAD FORECASTING

| Method | Forecasted peak load (*MW*) | | | Error criteria | | |
|---|---|---|---|---|---|---|
| | 2018 | 2019 | 2020 | MAE (*MW*) | MAPE (%) | RMSE (*MW*) |
| Real data | 55442 | 57098 | 57635 | --- | --- | --- |
| B-LSTM | 55597.5 | 57700.8 | 58218.3 | 447.2 | 0.788 | 492.54 |
| LSTM | 55829 | 58072 | 59693 | 1139.667 | 2 | 1333.39 |
| MLP | 53633.2 | 55237.5 | 56021.3 | 1761 | 3.10 | 1764.20 |

TABLE III
FORECASTED PEAK LOADS OF YEARS 2021-2028

| Year | 2021 | 2022 | 2023 | 2024 |
|---|---|---|---|---|
| Forecasted peak load (*MW*) | 58254 | 60259 | 62030 | 63555 |
| Year | 2025 | 2026 | 2027 | 2028 |
| Forecasted peak load (*MW*) | 64841 | 65906 | 66773 | 67470 |

The obtained results are presented in Table IV. The results of scheme IV are also illustrated in Fig. 7. Note that for WF and BES, the cumulative new installed capacity at each stage is presented. For example, in the first scheme, a WF with 72.95 MW capacity is installed in bus 6 at first stage, while in the second stage the installed capacity is increased to 200 MW. In scheme I, which is basically similar to the previous conventional models in the literature, a single-circuit and a double-circuit HVAC lines are constructed between buses 16-19 and 7-8 at the first stage. Two double-circuit HVAC lines between buses 3-25 and 14-26 are constructed at the third stage. At the last stage

TABLE IV
RESULTS OF THE PROPOSED HR CO-PLANNING MODEL

| | Investment Cost (IC) ($10^6\$$) | | | | | TOC ($10^6\$$) |
|---|---|---|---|---|---|---|
| Scheme | AC line | DC line | WF | BES | TIC | 4149.726 |
| | | | | | | TPC ($10^6\$$): |
| I | 552.348 | ---- | 747.141 | ---- | 1299.489 | Z=5449.215 |

**AC line:** (16-19), (7-8)* s=1, (14-26)*, (3-25)* s=3, (3-25)* s=4

**WF: Bus 6:** 72.95 MW, **Bus 23:** 200 MW s=1,
**Bus 6:** 200 MW, **Bus 9:** 175.75 MW s=2, **Bus 9:** 200 MW, **Bus 25:** 177.5 MW,
**Bus 26:** 118.065 MW s=3, **Bus 25:** 469.4 MW, **Bus 26:** 153.04 MW s=4

**Total wind curtailment:** 26610.549 MWh

| II | 128.283 | 191.64 | 747.141 | 313.48 | 1380.544 | 3716.429 Z=5096.973 |

**AC line:** (16-19), (7-8)* s=1,

**DC line:** (3-25), (14-26) s=3

**WF: Bus 6:** 72.95 MW, **Bus 23:** 200 MW s=1, **Bus 6:** 200 MW,
**Bus 9:** 175.75 MW s=2, **Bus 9:** 200 MW, **Bus 25:** 275.028 MW,
**Bus 26:** 166.498 MW s=3, **Bus 25:** 275.028 MW, **Bus 26:** 347.401 MW s=4

**BES: Bus 3:** 212.102 MW, 848.41 MWh,
**Bus 20:** 226.815 MW, 907.258 MWh s=1, **Bus 20:** 241.993 MW, 967.974 MWh,
**Bus 25:** 30.761 MW, 92.283 MWh, **Bus 26:** 129.822 MW, 517.905 MWh s=3,
**Bus 25:** 47.653 MW, 142.958 MWh s=4

**Total wind curtailment:** 1638.116 MWh

| III | 705.463 | ---- | 747.141 | ---- | 1452.604 | 4143.215 Z=5595.819 |

**AC:** (16-19), (7-8)* s=1, (3-25)*, (14-26)* s=3, 2×(19-20)*, (20-23)*, (3-25)* s=4

**WF: Bus 6:** 72.95 MW, **Bus 23:** 200 MW s=1, **Bus 6:** 200 MW,
**Bus 9:** 175.75 MW s=2, **Bus 9:** 200 MW, **Bus 25:** 177.5 MW,
**Bus 26:** 118.065 MW s=3, **Bus 25:** 459.118 MW, **Bus 26:** 163.311 MW s=4

**Total wind curtailment:** 26733.54 MWh

| IV | 243.774 | 252.931 | 747.141 | 317.386 | 1561.232 | 3697.553 Z=5258.785 |

**AC:** (6-10), (16-19), (7-8)* s=1, (19-20)* s=4

**DC:** (3-25), (14-26) s=3, (20-23) s=4

**WF: Bus 6:** 72.95 MW, **Bus 23:** 200 MW s=1, **Bus 6:** 200 MW,
**Bus 9:** 175.75 MW s=2, **Bus 9:** 200 MW, **Bus 25:** 124.07 MW,
**Bus 26:** 171.498 MW s=3, **Bus 25:** 308.265 MW, **Bus 26:** 314.16 MW s=4

**BES: Bus 3:** 217.81 MW, 869.503 MWh, **Bus 20:** 226.815 MW, 907.26 MWh s=1,
**Bus 20:** 237 MW, 947.62 MWh, **Bus 25:** 22.87 MW, 68.6 MWh,
**Bus 26:** 127 MW, 508 MWh s=3, **Bus 20:** 250 MW, 1000 MWh,
**Bus 25:** 73 MW, 219 MWh s=4

**Total wind curtailment:** 885.035 MWh

**\*:** Double Circuit Line.

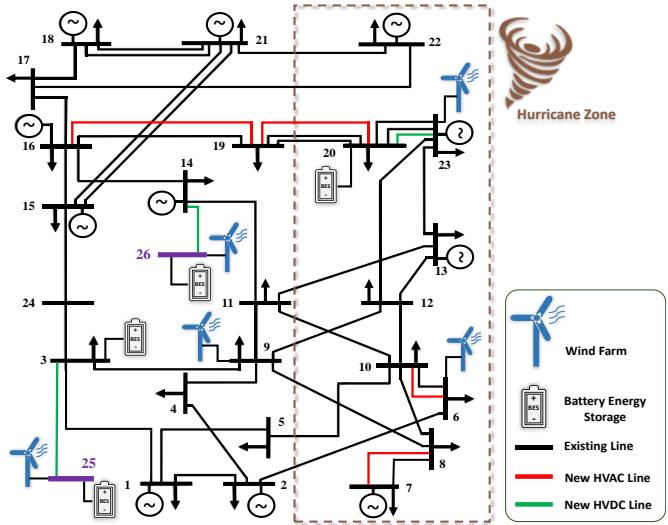

Fig. 7. The obtained results for scheme IV

another double-circuit HVAC line is constructed between buses 3-25. In scheme II, a single-circuit and a double-circuit HVAC lines are constructed between buses 16-19 and 7-8 at the first stage. Two HVDC lines between buses 3-25 and 14-26 are also constructed at the third stage of scheme II. In comparison to scheme I, the total planning cost of scheme II is saved up to 352.242 $10^6$\$. This cost saving confirms the impacts of HVDC lines and BES devices on minimizing the total costs of expansion co-planning problem. The impact of modeling power system resilience against probable hurricanes can be evaluated by comparing the schemes I and III, and also schemes II and IV. According to scheme II, by ignoring the system resilience against hurricanes, a total energy up to 309.058 GWh is interrupted with an expected cost of 971.745 $10^6$\$, while the TPC in scheme IV is just 161.812 $10^6$\$ more expensive than scheme II. Although the TIC in scheme IV is 180.688 $10^6$\$ more than scheme II, the TOC is saved due to more BES utilization, 753.081 MWh lower total wind curtailment, and new constructed HVAC/DC corridors for demand supplying. To enhance the system resilience, the cost of installed BES in scheme IV is 3.906 $10^6$\$ more than scheme II. Note that the equivalent WFs investment cost in all schemes, is due to similar RPS policy. The comparison of scheme III and IV, confirms the noticeable impact of HVDC lines and BESs in cost reduction under considering the power system resilience against hurricanes. In scheme III, 8 HVAC lines, including 7 double-circuit and one single-circuit lines, are constructed to enhance the system resilience. In this scheme, 26733.54 MWh wind power is curtailed due to lack of BESs. The wind curtailment is decreased to 885.035 MWh in scheme IV. In scheme IV using 4 HVAC, including 2 double-circuit and 2 single-circuit lines, and also 3 HVDC lines the system is able to withstand hurricanes with TPC saving of 337.034 $10^6$\$ related to the scheme III. As shown in Fig. 7, in scheme IV, 2 WFs, one BES, one HVDC, and 3 HVAC lines (2 double-circuit and one single-circuit lines) are constructed in the considered HZ. The buses 25 and 26, which are considered as remote WF buses, are connected to the system using 2 new HVDC lines. Also, 2 BESs are installed in buses 25 and 26 to handle the intermittency of WFs.

Generally, the proposed method in this paper is not case sensitive and can be applied over any system. It is expected that, for large-scale test systems, the computational time of the proposed method increases. However, the proposed BD technique facilitates the solution process under operational scenarios and resilience contingencies for large scale power systems. In fact, using the decomposed structure, the proposed model can be applied over large-scale systems providing that fast computational and parallel processing tools are utilized.

## VI. CONCLUSION

This paper presented a multi-stage hurricane resilient model for expansion co-planning of transmission lines, BESs, and WFs. To evaluate the system resilience against hurricanes, probable HS scenarios were generated using MCS. The major findings of this paper are summarized as follows. 1) The new planning tools, including the BES and HVDC options, result in a more economic planning scheme that facilitate the integration of remote WFs, especially in a HR co-planning model. 2) The traditional TEP models are not adequate to enable the power system to survive the HILP events. Also, using only traditional HVAC transmission lines to enhance the system resilience against hurricanes is not economical. 3) By additional investment cost the system resilience against HILP events can be achieved using HVAC, HVDC, WF, and BES planning tools. 4) The proposed CTPC approach can extract proper representative hours for each planning stage to capture the load and wind power uncertainties. Also, the B-LSMT networks are able to forecast the yearly load growth accurately.